\documentclass{article}
\usepackage{ijcai24}

\usepackage{times}
\usepackage{soul}
\usepackage{url}
\usepackage[hidelinks]{hyperref}
\usepackage[utf8]{inputenc}
\usepackage[small]{caption}
\usepackage{graphicx}
\usepackage{amsmath}
\usepackage{amsthm}
\usepackage{booktabs}
\usepackage{algorithm}
\usepackage{algorithmic}
\usepackage[switch]{lineno}

\usepackage{multirow}

\title{DKiS: Decay weight invertible image steganography with private key}

\author{
Hang Yang
\and
Yitian Xu\and
Xuhua Liu\\
\affiliations
College of Science,
China Agricultural University,
Beijing 100083, China\\
\emails
yanghang@cau.edu.cn,
xytshuxue@126.com,
liuxuhua@cau.edu.cn
}

\begin{document}

\maketitle

\begin{abstract}
Image steganography, defined as the practice of concealing information within another image, traditionally encounters security challenges when its methods become publicly known or are under attack. To address this, a novel private key-based image steganography technique has been introduced. This approach ensures the security of the hidden information, as access requires a corresponding private key, regardless of the public knowledge of the steganography method. Experimental evidence has been presented, demonstrating the effectiveness of our method and showcasing its real-world applicability. Furthermore, a critical challenge in the invertible image steganography process has been identified by us: the transfer of non-essential, or `garbage', information from the secret to the host pipeline. To tackle this issue, the decay weight has been introduced to control the information transfer, effectively filtering out irrelevant data and enhancing the performance of image steganography. The code for this technique is publicly accessible at https://anonymous.4open.science/r/DKiS/README.md, and a practical demonstration can be found at http://47.94.105.69/hidekey/.
\end{abstract}

\section{Introduction}

Image steganography, a subset of the broader field of steganography, involves the practice of concealing information within a digital image \cite{Kessler2011Overview,Johnson1998Exploring}. This technique is rooted in the ancient art of hidden communication, but has evolved significantly with the advent of digital technologies. At its core, image steganography aims to embed data, such as text, audio, or another image, into a host image in a manner that is undetectable to the casual observer \cite{RIIS}. The primary objective is to maintain the apparent normalcy of the host image while secretly carrying additional information.

\begin{figure}[!h]
	\centering
	\includegraphics[width=0.45\textwidth]{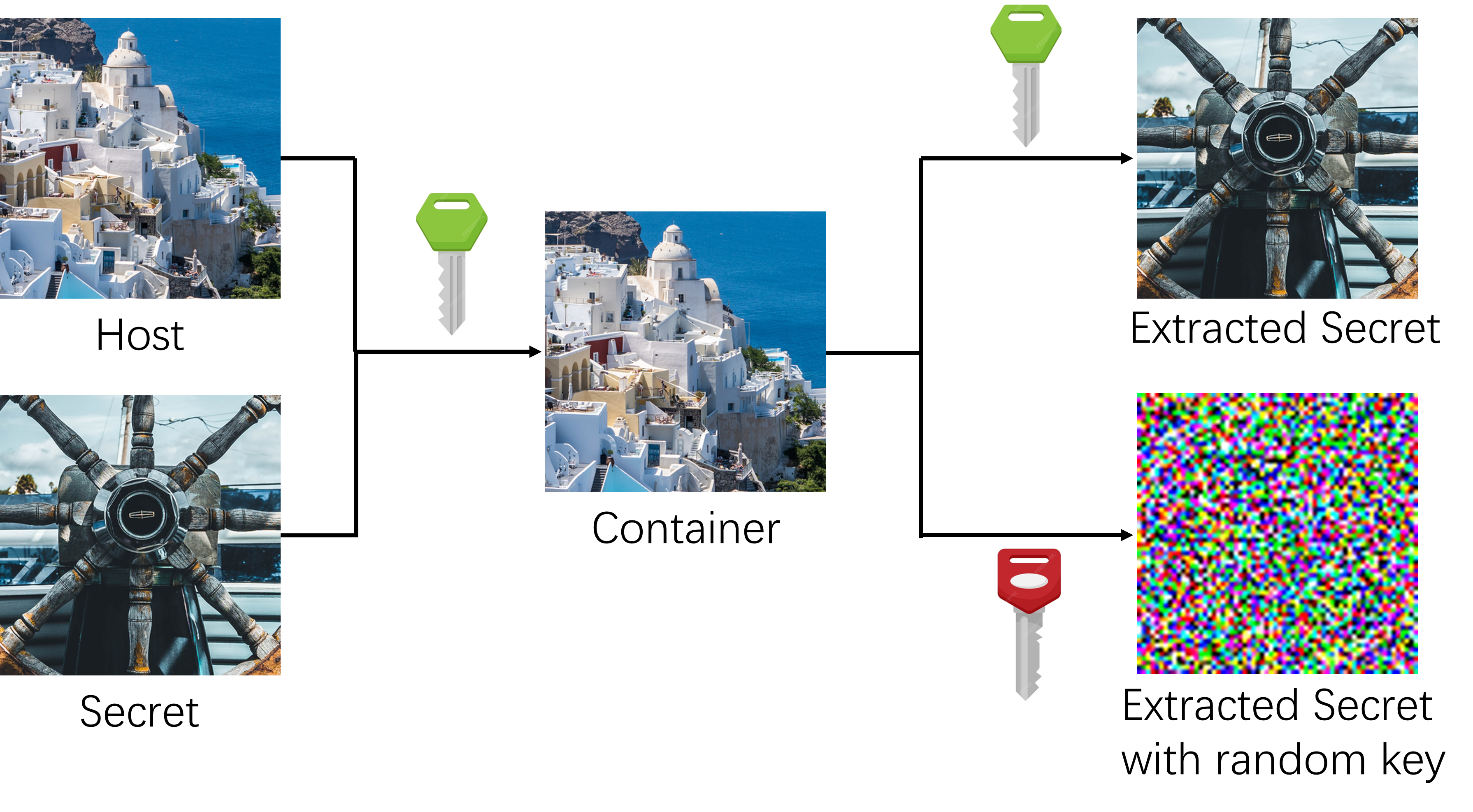}
	\caption{The workflow of image steganography with a private key.}
	\label{embedandextract}
\end{figure}

The evolution of image steganography has been marked by various methods and technologies. Initially, simple techniques \cite{luo2010edge,chan2004hiding} like the Least Significant Bit (LSB) insertion were used, where information is embedded in the least significant bits of the pixel values of an image, causing minimal visual alteration \cite{kadhim2019comprehensive}. Some other methods \cite{barni2001improved,hsu1999hidden} include transform domain techniques, where information is embedded in frequency domain of an image, making it more resilient to compression and other image processing operations.

The introduction of deep learning has further advanced the field of image steganography \cite{lu2021large,zhang2020udh,baluja2019hiding,duan2019reversible,duan2020high,duan2020highx}. By employing complex neural networks, it has become possible to embed larger amounts of data with improved security measures. These deep learning models can learn optimal ways to conceal data within images, making detection by third-party observers or automated systems significantly more challenging.

Image steganography continues to evolve, facing the challenge of maintaining confidentiality amidst widespread popularity. The known methods of steganography can be compromised by unauthorized parties, making it unsuitable for high-security contexts. To address this, our research integrates preset private keys into high-capacity image steganography, as shown in Fig. \ref{embedandextract}, based on invertible neural networks. This innovative approach secures the embedded data even if the steganographic technique is widely known, using preset private keys that eliminate the need for transmission with the container image—thereby reducing interception risks. Our method, which combines advanced deep learning algorithms with this novel private key strategy, shows promising effectiveness in our experimental results. This highlights its potential for various applications, including secure communications and confidential data storage, where stringent security measures are paramount.

In addition, we find that when the information passing through the secret pipeline, the amount of secret part is getting less and less, which gives us the idea to limit the information transfer from secret pipeline into the host pipeline. Therefore we proposed the decay weight for a better performance.

The main contributions are listed as follows:

1. Preset private key is introduced into deep learning based image steganography for the first time, which extremely increase the security of image steganography.

2. Decay weight is proposed to control the amount of information transfer into the host pipeline from the secret pipeline based on the insight that the secret information is getting less and less as the secret image pass through the secret pipeline.

\section{Related Works}
\subsection{Image Steganography without Private Key}
Image steganography is broadly categorized into traditional and deep learning-based methods.

\textbf{Traditional Image Steganography:}
The most prevalent method in this category is the Least Significant Bit (LSB) technique \cite{chan2004hiding,tamimi2013hiding}, which embeds secret messages by altering the $n$ least significant bits of the host image. However, these modifications make it susceptible to detection by steganalysis methods \cite{fridrich2001detecting,hawi2004steganalysis,zhi2003lsb}. Other notable techniques include Pan's pixel value differencing (PVD) \cite{pan2011image}, Tsai's histogram shifting \cite{tsai2009reversible}, Nguyen's use of multiple-bit-planes \cite{nguyen2006multi}, and Imaizumi's palette-based approach \cite{imaizumi2014multibit}. Some traditional methods utilize frequency domains like discrete cosine transform (DCT) \cite{hsu1999hidden}, discrete Fourier transform (DFT) \cite{ruanaidh1996phase}, and discrete wavelet transform (DWT) \cite{barni2001improved}. These methods offer more robustness and stealth than LSB techniques, though they still face limitations in payload capacity.

\textbf{Deep Learning-Based Image Steganography:}
Recent advancements have seen deep learning techniques significantly outperform traditional methods in image steganography. HiDDeN \cite{zhu2018hidden} and SteganoGAN \cite{zhang2019steganogan} utilize an encoder-decoder architecture, coupled with a third network to counteract steganalysis. Shi’s Ssgan \cite{shi2018ssgan} is based on generative adversarial networks. Baluja \cite{Baluja2017Hiding,baluja2019hiding} and Zhang \cite{zhang2020udh} have achieved higher payload capacities by hiding secret images of the same size as the host image. Lu \cite{lu2021large}, Jing \cite{jing2021hinet}, and Jia \cite{jia2023afcihnet} have introduced invertible neural networks (INN) into the field and achieved SOTA performance in this area.

\subsection{Image Steganography with Private Key}
To address security issues arising from the widespread use of steganographic methods, researchers have proposed LSB-based image steganography utilizing private keys \cite{karim2011new,almazaydeh2018image,al2019secret}. While this approach enhances security, it tends to suffer from low payload capacity. Seeking to increase capacity, Kweon \cite{kweon2021deep} integrated private key into a deep learning-based steganography framework. Although this method achieves higher capacities, the private key is generated with the container image rather than preset. Therefore the transmission of the private key with the container image is required, which could potentially compromise security. Additionally, Kweon's method faces low performance issues. The quality of generated image is not satisfied.

\section{Method}
\subsection{Overview}
\begin{figure}[!h]
	\centering
	\includegraphics[width=0.45\textwidth]{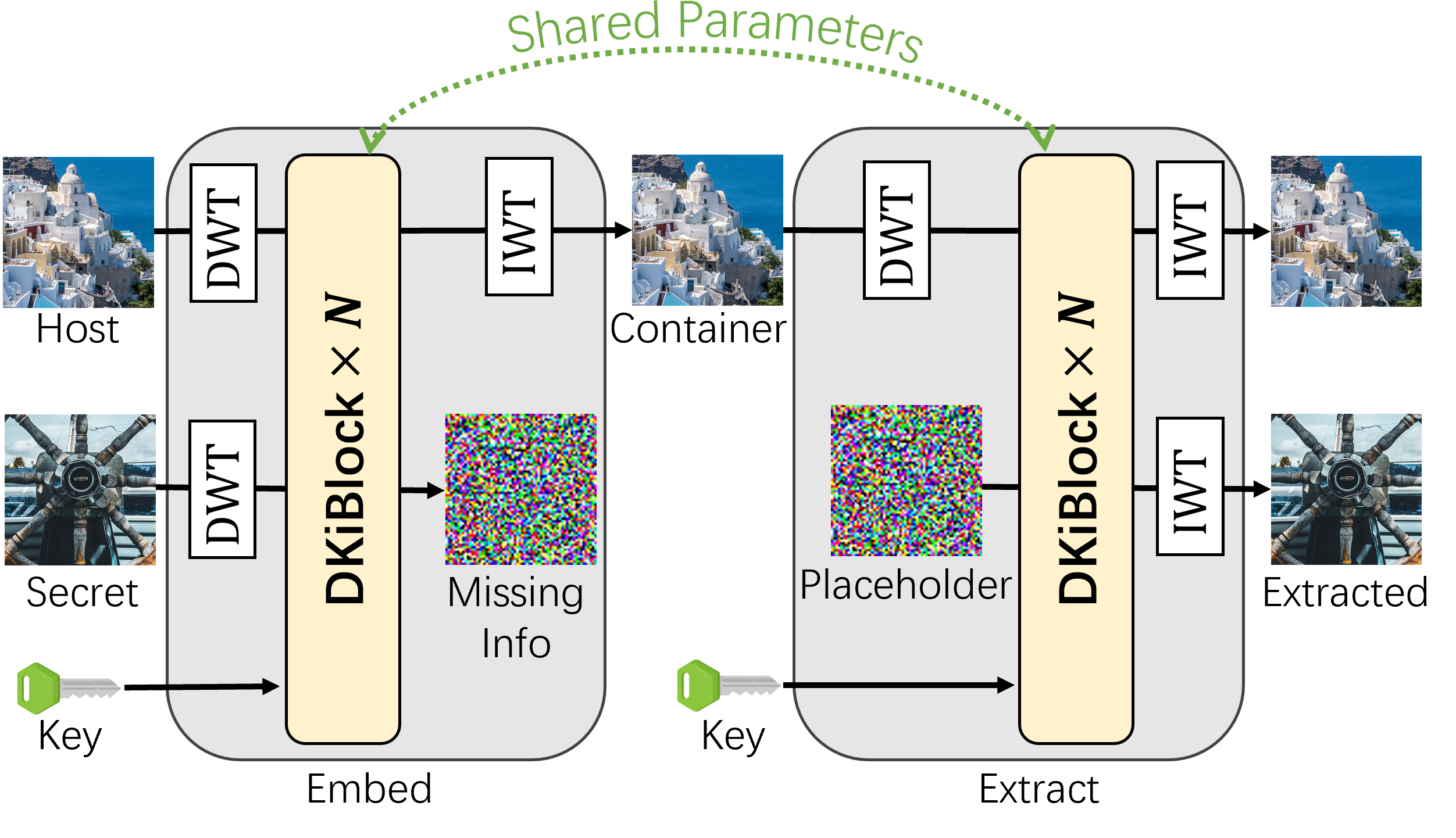}
	\caption{The overview of DKiS.}
	\label{DKiS}
\end{figure}

The architecture of our DKiS model is illustrated in Fig. \ref{DKiS}. In the embedding phase, the process involves the host image, secret image, and private key as inputs, producing the container image and what we term as `missing info'. The term `missing info' is used because, during transmission, only the container image is sent to the receiver, resulting in the missing of the other image. DKiS requires two images for both embedding and extraction phases. Consequently, a placeholder Gaussian distribution is introduced into DKiS alongside the container image and key. And the generated extracted image is indistinguishable from the original secret image if and only if the key used matches the one employed during the embedding phase.

Additionally, before the host, secret, and container images are processed by the DKiBlocks, a discrete wavelet transform (DWT) is applied to these input images. Correspondingly, an inverse discrete wavelet transform (IWT) is applied to the output from the DKiBlocks based on the work of \cite{jing2021hinet}.

\subsection{DKiBlock}
\begin{figure*}[!h]
	\centering
	\includegraphics[width=0.85\textwidth]{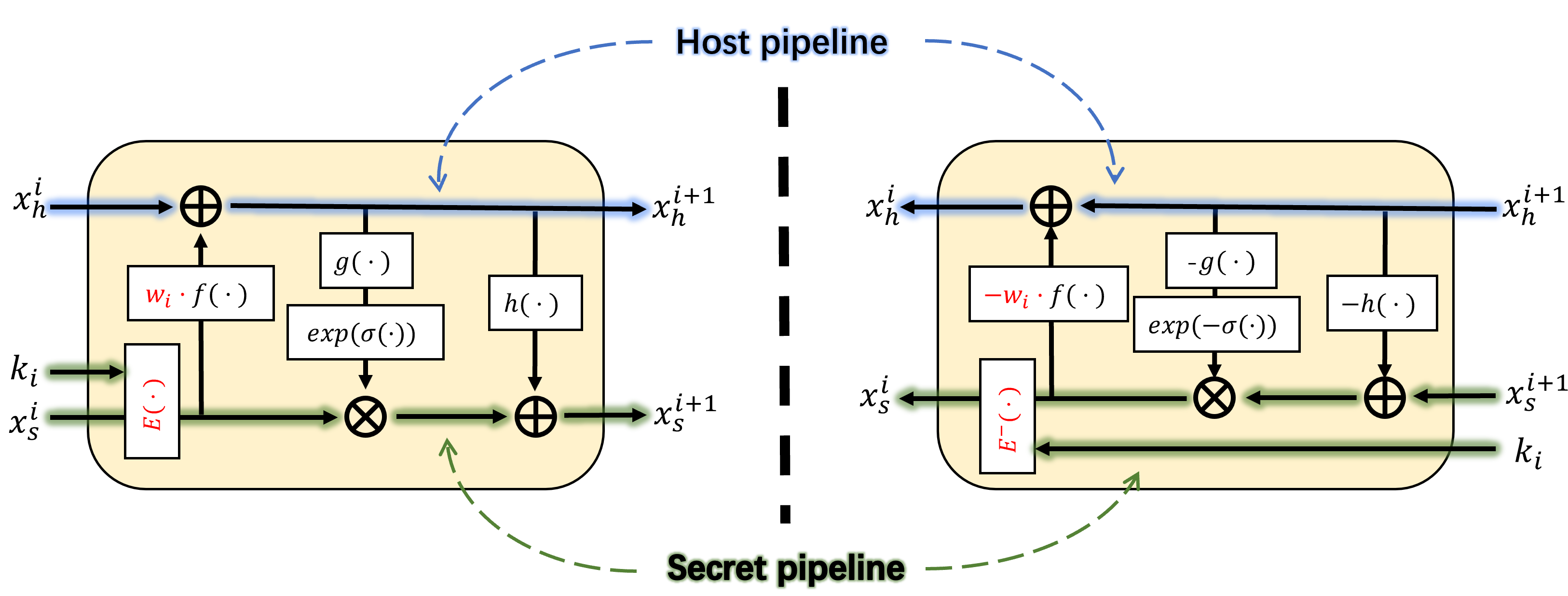}
	\caption{The architecture of DKiBlock. Left: forward process. Right: inverse process.}
	\label{DKiBlock}
\end{figure*}

The architecture of the DKiBlock is depicted in Fig. \ref{DKiBlock}. This design is founded on invertible blocks \cite{xiao2020invertible}, enhanced through the incorporation of decay weights and private key.

The forward and inverse processes are shown as Eqs. \eqref{ibf2} and \eqref{ibb2} respectively.

\begin{align}
	x_k^{i+1} &= E(x_s^i, k_i) \nonumber \\
	x_h^{i+1} &= x_h^i + w_i\cdot f(x_k^i)  \label{ibf2} \\	
	x_s^{i+1} &= x_k^i \otimes exp(\sigma(g(x_h^{i+1}))) + h(x_h^{i+1}) \nonumber 
\end{align}
With a little transformation of Eq. \eqref{ibf2}, the inverse processes is gained and shown as below:
\begin{align}
	x_k^i &= (x_s^{i+1}-h(x_h^{i+1})) \otimes exp(-\sigma(g(x_h^{i+1}))) \nonumber   \\
	x_h^i &= x_h^{i+1}-w\cdot f(x_k^{i})\label{ibb2} \\
	x_s^i &= E^-(x_k^{i+1}, k_i) \nonumber 
\end{align}
Where $E$ is the encoding function, it encodes the input $x$ with $k$, and in order to maintain the invertible ability, it needs to have invert function $E^-$, the detail of $E$ will be discussed in Section 3.4. And $w_i$ is the decay weight, the detail and purpose will be discussed in Section 3.3.

\subsection{Decay Weight}
The extraction process becomes more challenging as the amount of secret information in `mission info' increases. For optimal performance, it is crucial to minimize the information about the secret image in the `missing info'. Ideally, this `missing info' should closely resemble the placeholder Gaussian distribution.

\begin{figure*}[!h]
	\centering
	\includegraphics[width=0.9\textwidth]{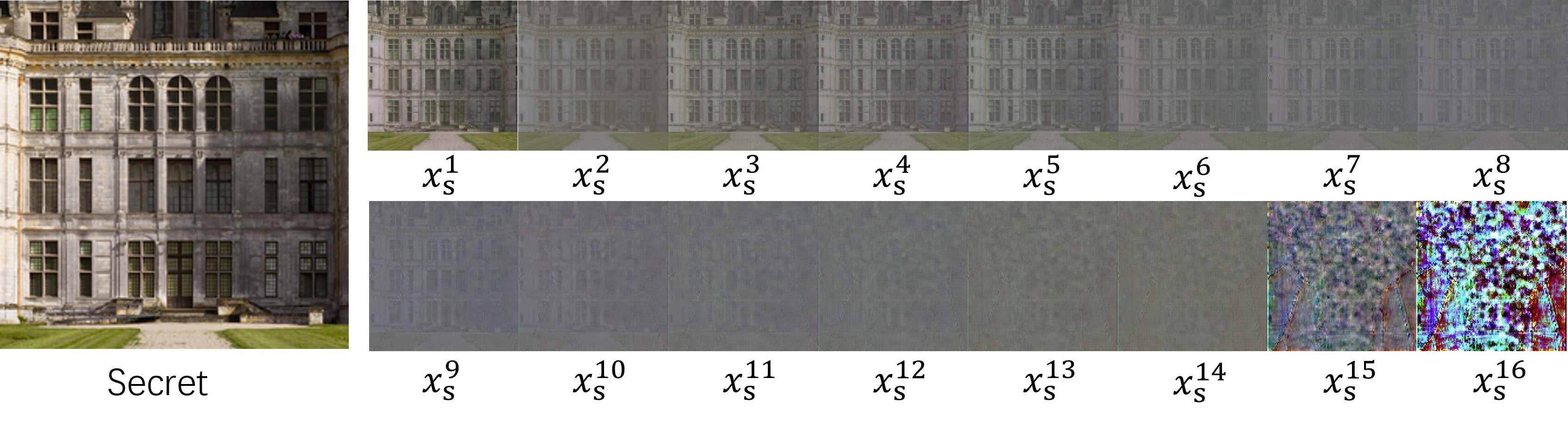}
	\caption{The outputs of secret pipeline.}
	\label{xs}
\end{figure*}
\begin{figure}[!h]
	\centering
	\includegraphics[width=0.45\textwidth]{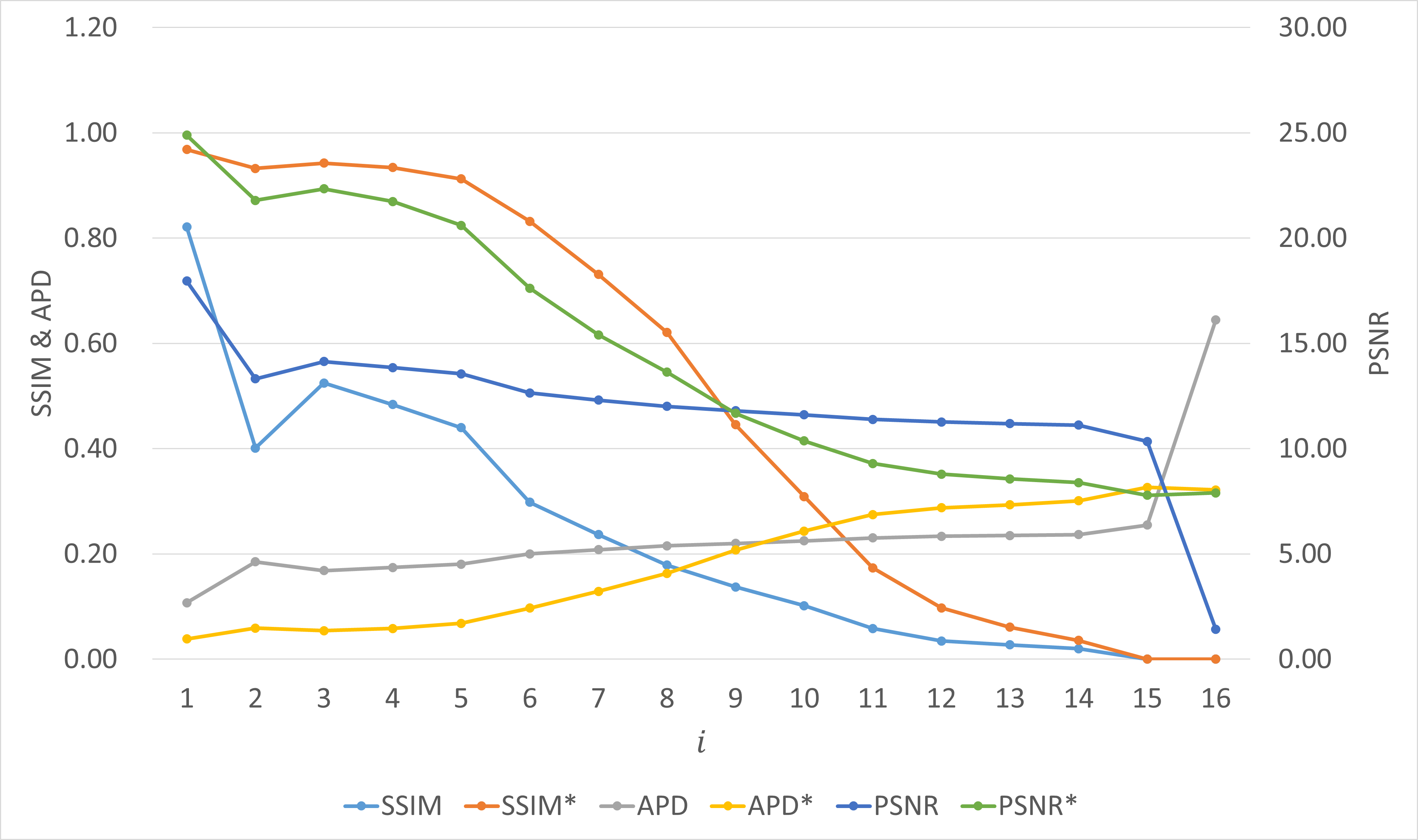}
	\caption{The SSIM, APD and PSNR values between each $x_s^i$ and the secret image. * denotes that $x_s^i$ has been aligned to the secret image through a linear transformation. This alignment ensures that $x_s^i$ matches the secret image in terms of mean and standard deviation. This figure shows that the similarity/distance between $x_s^i$ and secret image is getting less/more as $i$ increasing.}
	\label{xsf}
\end{figure}
However, the transformation from secret image to `missing info' is impossible to finish in one invertible block. To better understand this, we analyzed the outputs of the secret pipeline, as depicted in Fig. \ref{xs}. We observed that the deviation of each $x_s^i$ from the secret image increases with the increment of $i$. To quantify these deviations, we calculated the distances/similarity between the secret image and each $x_s^i$, employing metrics such as Average Pixel Distance (APD), Peak Signal-to-Noise Ratio (PSNR), and the Structural Similarity Index Model (SSIM), which are elaborated in Section 4.2. The comparative findings are displayed in Fig. \ref{xsf}. This analysis indicates a progressive divergence of $x_s^i$ from the secret image with increasing $i$, signifying a reduction in the information content of the secret image as it progresses through the secret pipeline.

Recognizing that less secret image information in $x_s^i$ reduces the necessity for the host pipeline, we introduced a decay weight $w_i$ to control the amount of information transferred from the secret to the host pipeline. The formulation of $w_i$ is detailed in Eq. \ref{w}, where $r$ represents the decay rate.
\begin{equation}
	\label{w}
	w_i = r^i
\end{equation}

The decay rate $r$ is between 0 to 1. The value of decay weight is getting less and less with increasing $i$, which reduces the information transfer from secret to host pipeline right along the $x_s^i$ becoming useless to the host pipeline.

\subsection{Private Key}
\begin{figure}[!h]
	\centering
	\includegraphics[width=0.45\textwidth]{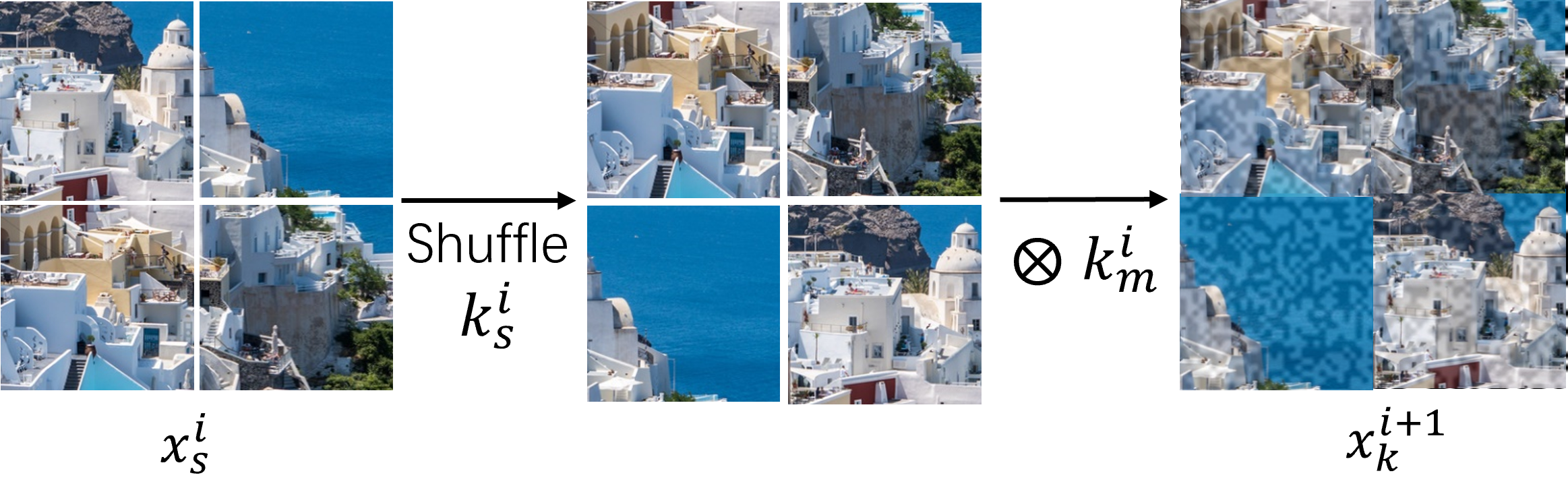}
	\caption{The overview of encode operation.}
	\label{encode}
\end{figure}

We have incorporated the private key into the secret pipeline through an encoding function $E(\cdot)$, as illustrated in Fig. \ref{encode}. 

This encoding process comprises two main steps: shuffling and element-wise multiplication. The private key, $k_i$, is divided into two components: $k_s^i$ and $k_m^i$. In the first step, the image is segmented into $4 \times 4$ small patches (Fig. \ref{encode} only displays $2 \times 2$ patches for clarity reason). These patches are then shuffled according to the sequence determined by $k_s^i$. In the second step, the image undergoes element-wise multiplication by $k_m^i$. Here, $k_m^i$ is of the same dimensions as $x_s^i$, and each element in $k_m^i$ has an equal probability of being either -1 or 1. It is important to note that the encoding operation is invertible solely when the private key is known.

\textbf{Key Generation Process:} To generate the key, any user-preset key is first transformed into a 256-bit number using SHA-256 encoding \cite{lilly2004device}. This encoded number is then converted into a random seed, which is subsequently used to generate both $k_s$ and $k_m$.

\subsection{Loss Function}

Our study cares two distinct types of similarity. Firstly, we examine the differences between the host image $x_h$ and the container image $x_c$, referred to as the C-pair. Secondly, we consider the differences between the secret images $x_s$ and the extracted images $x_e$, termed the S-pair. To quantify these differences, we introduce two specific loss functions, $L_c$ and $L_s$, defined as follows:
\begin{align}
	L_c= \sum_{p}(x_c^{(p)}-x_h^{(p)})^2, \quad L_s= \sum_{p}(x_s^{(p)} - x_e^{(p)})^2
\end{align}
In these equations, $x^{(p)}$ denotes the pixel value at position $p$ in image $x$. The terms $L_c$ and $L_s$ are designed to measure the distant within the C-pair and S-pair, respectively. The overall loss function is then computed as a weighted sum of these two losses:
\begin{equation}
	L = \lambda_cL_c + \lambda_sL_s
\end{equation}
where $\lambda_c$ and $\lambda_s$ are weighting factors for the respective loss functions.

\section{Experiment}
\subsection{Implementation Details}
Unless otherwise specified, our model is trained on the DIV2K dataset \cite{agustsson2017ntire}. We pre-process the input images by cropping them to a resolution of 256 $\times$ 256 pixels. To enhance generalization, we employ random cropping on the training dataset. Conversely, for the testing dataset, center cropping is utilized to eliminate randomness in evaluation. The weighting factors $\lambda_c$ and $\lambda_s$ are both set to 1. The training process involves 1600 epochs, utilizing the Adam optimizer with a learning rate of $10^{-4.5}$, and $\beta_1=0.9$, $\beta_2=0.99$. The learning rate is halved every 200 epochs to optimize convergence.

Acknowledging the inevitability of rounding errors in real-world applications, we incorporate a rounding operation on the container image in all our experiments to simulate this effect. To address the non-differentiable nature of the rounding operation, we employ the Gradient Approximation Function (GAF) as described in PRIS \cite{yang2023pris}.

\subsection{Evaluation Criteria}
The primary goal of image steganography is to maximize similarity in the C-pair (container and host images) and the S-pair (secret and extracted images). Introducing a private key into our system, we define a new pair called the S'-pair, which consists of the secret image and an incorrectly extracted secret image using an incorrect key. 

Our evaluation employs the Peak Signal-to-Noise Ratio (PSNR) and the Structural Similarity Index Model (SSIM) \cite{setiadi2021psnr} to measure image similarity, with higher values indicating greater similarity. These metrics are denoted as PSNR-x and SSIM-x, where `x' specifies the image pair under evaluation. For C-pair and S-pair, higher PSNR and SSIM values suggest superior steganographic performance. In contrast, lower PSNR and SSIM values for the S'-pair indicate enhanced security against unauthorized information extraction.

Additionally, we utilize the Average Pixel Distance (APD) for quantifying the pixel-wise distance between two images, as defined in \ref{apd}. In this context, $x^{(p)}$ and $y^{(p)}$ represent pixel values of images $x$ and $y$ at position $p$, respectively, and $N$ is the total number of pixels. A higher APD value implies a greater distance between the images.
\begin{equation}
	APD(x, y) = \frac{1}{N}\sum_p|x^{(p)}-y^{(p)}|
	\label{apd}
\end{equation}

\subsection{Ablation Study}
To illustrate the effectiveness of our proposed decay weights and standardize pre-process techniques, we conducted ablation studies. As shown in Table \ref{ablation}, the incorporation of decay weight and standardize pre-process with $r=0.6$ resulted in a significant improvement, reducing the loss value by 23.14\%.
\begin{table*}[!h]
	\centering
	\caption{Ablation studies for normal pre-process and decay rate.}
	\begin{tabular}{ccccccc}
		\toprule
		Pre-process & Decay rate &  PSNR-C$\uparrow$ & PSNR-S$\uparrow$ & SSIM-C$\uparrow$  & SSIM-S$\uparrow$ & Loss$\downarrow$            \\ \midrule
		Normalize & $\times$ & 40.63 & 40.39 & .9928 & .9987  & 42.65 \\
		Standardize & $\times$ & 41.40  & 41.11 & .9945 & .9989 & 35.41   \\
		Standardize & 0.9 & 41.25  & 41.21 & .9945 & .9989 & 35.77   \\
		Standardize & 0.8 & 41.49  & 41.21 & .9947 & .9989 & 34.88   \\
		Standardize & 0.7 & 41.73  & 41.43 & .9946 & .9990 &  32.78  \\
		Standardize & 0.6 &  \textbf{41.91} & \textbf{41.62} & \textbf{.9948} & \textbf{.9990} &  \textbf{31.48}  \\
		Standardize & 0.5 &  41.54 & 40.97 & .9943 & .9989 &  36.05  \\
		\bottomrule
		\label{ablation}
	\end{tabular}
\end{table*}

\begin{figure}[!h]
	\centering
	\includegraphics[width=0.45\textwidth]{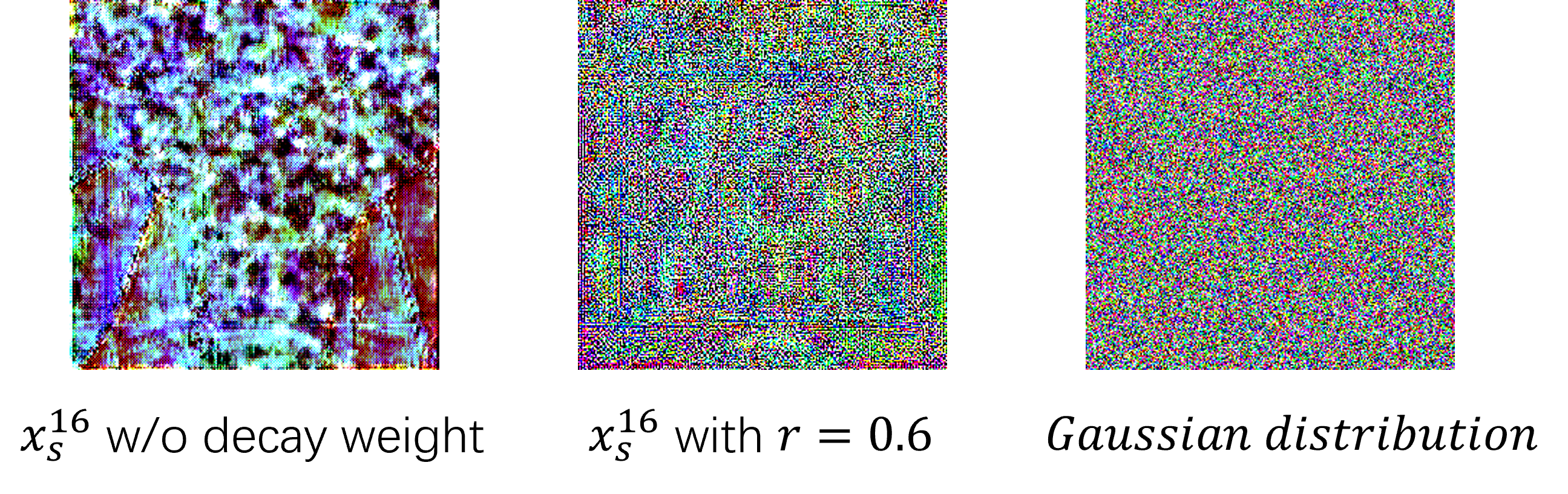}
	\caption{The comparison of the missing info of secret pipeline between without decay weight, with $r=0.6$ and Gaussian distribution.}
	\label{xs16}
\end{figure}

 Furthermore, We conducted analyses to evaluate the effects of implementing decay weights on the missing information in our model. These comparisons included scenarios without decay weight, with a decay weight characterized by $r=0.6$, and with a Gaussian distribution. The results, as illustrated in Fig. \ref{xs16}, reveal that the introduction of the decay weight leads to a pattern of missing information that more closely aligns with that of the Gaussian distribution. This alignment provides insight into why the decay weight is effective in our model.

\subsection{Experiments with Private Key}
\begin{figure}[!h]
	\centering
	\includegraphics[width=0.49\textwidth]{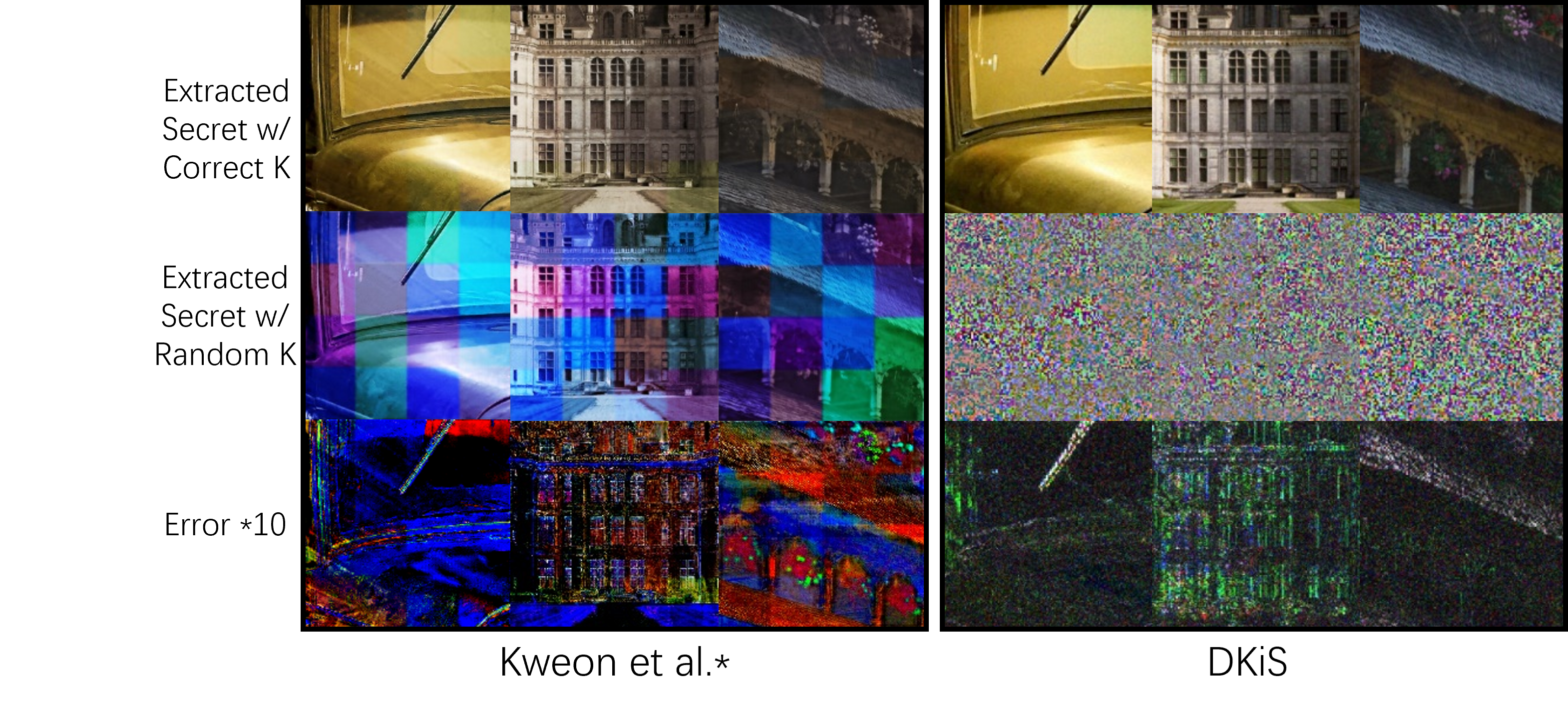}
	\caption{Visual comparison.}
	\label{vcompare}
\end{figure}

Due to the novelty of our method, which involves hiding an image within another image using a private key, direct comparisons with existing techniques are challenging. The most closely related work \cite{kweon2021deep} operates under a significant constraint: the key is generated by embedding networks, and cannot be preset. This implies that the key must be transmitted alongside the container image, a limitation our method overcomes. Consequently, a direct, fair comparison with existing methods is not feasible. Despite the increased practicality and complexity of our task compared to previous work, our approach demonstrates superior performance, as evidenced by the results presented in Tab. \ref{compare} and Fig. \ref{vcompare}.
\begin{table*}[!h]
	\centering
	\caption{Results with private key.}
	\begin{tabular}{ccccccc}
		\toprule
		Method &  PSNR-C$\uparrow$ & PSNR-S$\uparrow$ & SSIM-C$\uparrow$  & SSIM-S$\uparrow$ & PSNR-S'$\downarrow$ & SSIM-S'$\downarrow$           \\ \midrule
		Kweon \cite{kweon2021deep} & 22.73 & 26.39 & .9485 & .9859  &11.52 & .4109 \\
		Our & 36.04 & 31.95 & .9858 & .9906 & 8.71 & .0014 \\
		\bottomrule
		\label{compare}
	\end{tabular}
\end{table*}

We expanded our experimentation to include additional datasets, as summarized in Tab. \ref{dataset}. We incorporated a subset of the COCO-2017 dataset \cite{cocodataset}, selecting 6,000 images at random, to evaluate our model's performance on diverse real-world scenarios. Additionally, we used a set of 1,000 images from the ImageNet-2012 \cite{ILSVRC15} to further assess the model's robustness in varied photographic contexts. Furthermore, to test the model's applicability in a different domain, we employed a sample of 1,000 images from PubLayNet \cite{zhong2019publaynet}, a dataset only consisting of document images, contrasting the real-world photos in the other datasets. The consistent performance across these varied datasets underscores the versatility and adaptability of our method in handling both photographic and document imagery.
\begin{table*}[!h]
	\centering
	\caption{Experiment results on other datasets.}
	\begin{tabular}{ccccccc}
		\toprule
		Dataset &  PSNR-C$\uparrow$ & PSNR-S$\uparrow$ & SSIM-C$\uparrow$  & SSIM-S$\uparrow$ & PSNR-S'$\downarrow$ & SSIM-S'$\downarrow$           \\ \midrule
		DIV & 36.04 & 31.95 & .9858 & .9906 & 8.66 & .0000 \\
		COCO & 32.92 & 30.33 & .9938 & .9886  &8.49 & .0000\\
		ImageNet & 33.25 & 31.01 & .9950 & .9900 & 8.32 & .0002 \\
		PubLayNet & 36.88 & 27.25 & .9969 &.9764 & 10.84 & .0000 \\
		\bottomrule
		\label{dataset}
	\end{tabular}
\end{table*}

\subsection{How is the secret image hiding?}
\begin{figure}[!h]
	\centering
	\includegraphics[width=0.45\textwidth]{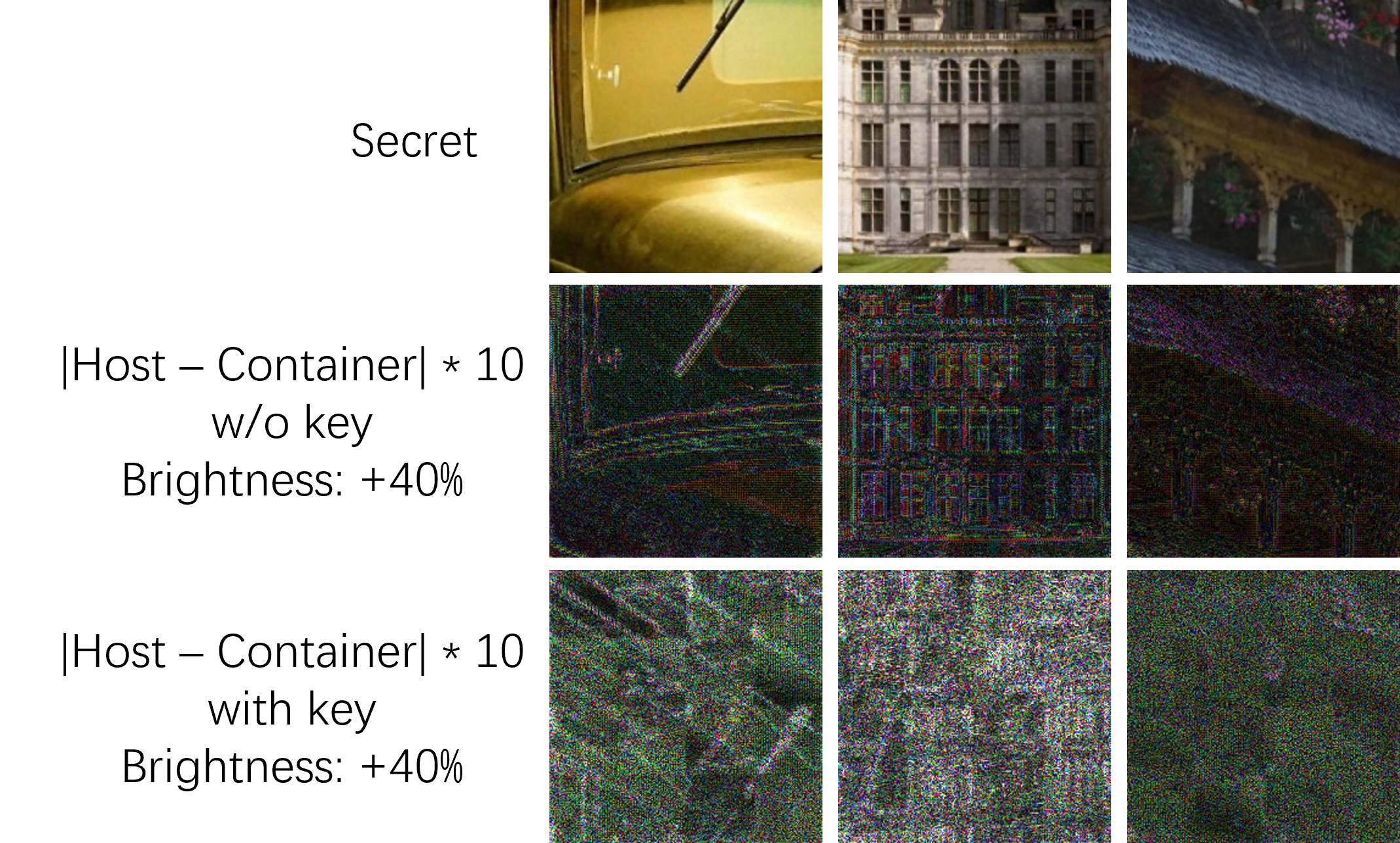}
	\caption{The difference between host and container images. }
	\label{dh}
\end{figure}

To illustrate how the secret image is concealed, we analyzed the differences between host and container images, as depicted in Fig. \ref{dh}. The first row displays the secret images. The second and third rows show the differences between C-pair without and with the private key, respectively. Since these differences are typically visually undetectable, we enhanced the images in the last two rows for clearer visualization. This enhancement involved multiplying the differences by a factor of 10 and increasing the brightness by 40\%.

The first two rows exhibit similar patterns, suggesting that the secret image is embedded in the host image through a specific addition operation. However, with the introduction of a private key, these shared patterns are no longer evident. This indicates that the private key modifies the embedding process  with a more complicated  way, thereby enhancing the security of the embedded image.

\subsection{Attack Simulation}
Since our main focus is the security of our method, and image steganography faces potential threats from various attacks, making it crucial to consider these in a real-world context. Attack could happen both on embedding and extraction phases in practical.

\textbf{\textit{Attack on Embedding:}} This attack involves unauthorized embedding of a secret image into a host image. The attacking network processes a host image alongside a fabricated secret, generating a deceptive container image. The critical measure of this attack's success lies in whether the target image steganography technique can inadvertently extract this false secret from the manipulated container image.

\textbf{\textit{Attack on Extraction:}} Here, the attack focuses on illegally extracting the secret image from the container image, bypassing the target image steganography method.

To rigorously evaluate our DKiS's security after implementing a private key, we designed two attack simulation networks based on DKiS and Resnet34 models. For the embedding attack, we trained the simulation networks with host and secret images as inputs and the container image as the output. Conversely, for the extraction attack, the container image was the input and the secret image was the output.

\begin{figure}[!h]
	\centering
	\includegraphics[width=0.49\textwidth]{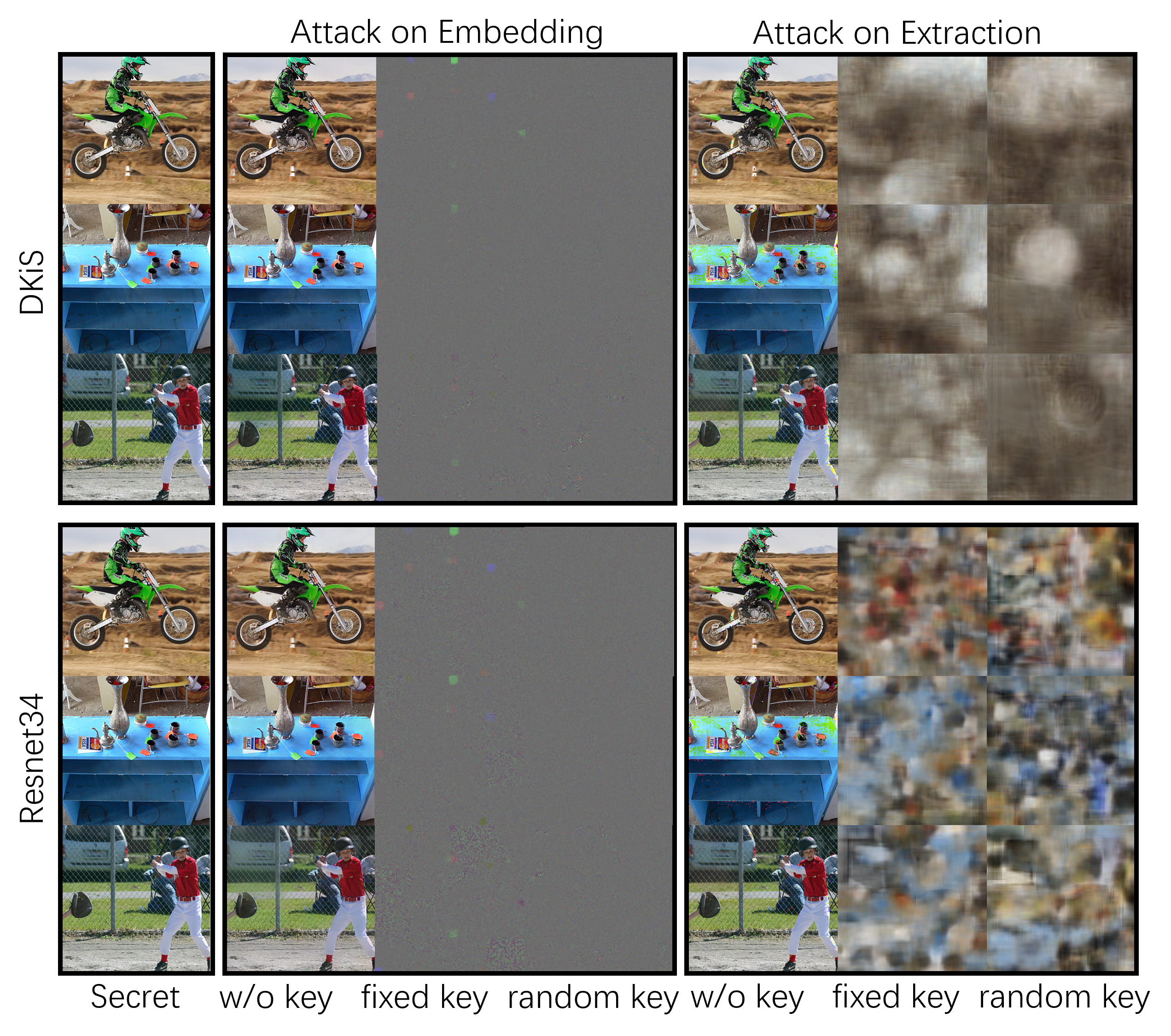}
	\caption{The extracted image in attacks.}
	\label{attackfig}
\end{figure}

The results, presented in Table \ref{attack}, illustrate that the introduction of the private key significantly impedes the attack simulation network's ability to hack the embedding and extraction processes. As further elucidated in Fig. \ref{attackfig}, while the attack networks were initially able to manipulate the secret image, the implementation of the private key rendered these attempts unsuccessful. This outcome affirms the enhanced security of our DKiS. Moreover, these results indicate that, even when the image steganography method is not public, the absence of a private key poses a significant security challenge, underscoring the necessity of the private key.

\begin{table*}[!h]
	\centering
	\caption{The PSNR and SSIM values between secret and extracted secret image under attacks.}
	\label{attack}
	\begin{tabular}{c|c|cc|cc}
		\hline
		\multirow{2}{*}{\begin{tabular}[c]{@{}c@{}}Attack\\ Model\end{tabular}} & \multirow{2}{*}{key} & \multicolumn{2}{c|}{Embedding}          & \multicolumn{2}{c}{Extraction}         \\ \cline{3-6} 
		&                      & \multicolumn{1}{c|}{PSNR$\downarrow$}  & SSIM$\downarrow$   & \multicolumn{1}{c|}{PSNR$\downarrow$}  & SSIM$\downarrow$   \\ \hline
		DKiS                                                                    & $\times$                  & \multicolumn{1}{c|}{34.50} & .9952 & \multicolumn{1}{c|}{35.08} & .9962 \\
		DKiS                                                                    & fixed                & \multicolumn{1}{c|}{11.40} & .0020 & \multicolumn{1}{c|}{12.43} & .2615 \\
		DKiS                                                                    & random               & \multicolumn{1}{c|}{11.39} & .0000 & \multicolumn{1}{c|}{10.67} & .0374 \\
		Resnet34                                                                &$\times$                  & \multicolumn{1}{c|}{29.63} & .9881 & \multicolumn{1}{c|}{35.45} & .9964 \\
		Resnet34                                                                & fixed                & \multicolumn{1}{c|}{11.38} & .0028 & \multicolumn{1}{c|}{10.47} & .0280 \\
		Resnet34                                                                & random               & \multicolumn{1}{c|}{11.39}     & .0000      & \multicolumn{1}{c|}{10.14} & .0313 \\ \hline
	\end{tabular}
\end{table*}

\section{Application}
\begin{figure}[!h]
	\centering
	\includegraphics[width=0.45\textwidth]{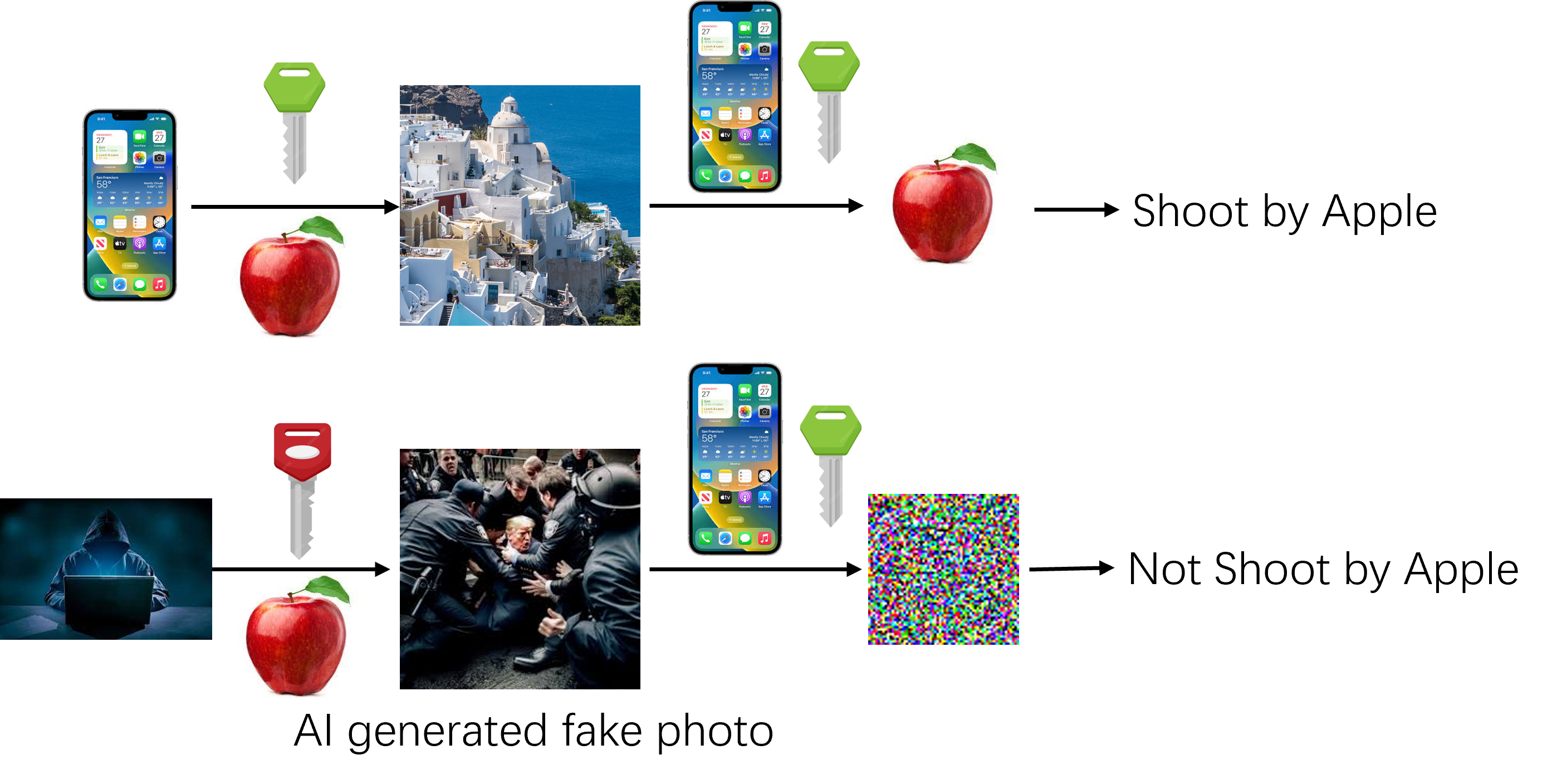}
	\caption{Photo verification.}
	\label{applicant}
\end{figure}

The integration of a private key significantly enhances the security of image hiding techniques, particularly for methods that are publicly accessible or likely attacked. For example, the use of the private key enables our method to address challenges that traditional image hiding approaches cannot solve. For instance, consider the application in photo verification. As illustrated in Fig. \ref{applicant}, camera manufacturers can embed a secret verification image into photos using a private key known only to them. If verification is required, users can submit the photo to the manufacturer. Successful extraction of the secret verification image by the manufacturer will confirm that the photo was taken with their product and has not been modified. This approach could become a universal standard among camera manufacturers, with each setting their unique private key. In the event of a key leaking, the impact would be limited to the compromised manufacturer, while the integrity of other cameras remains unaffected. 

The application of DKiS in differentiating real photos/videos from AI-generated content is especially pertinent given the rapid advancements in AI imagery. Its capability to embed secret watermarks in sensitive and critical media, such as CCTV footage or press photographs, enhances security and ensures the authenticity of these sources. In an era where digital authenticity is increasingly challenged, DKiS offers a much-needed solution to uphold the integrity of digital media.

\section{Conclusion}
In this paper, we introduced DKiS, a pioneering approach in the realm of image steganography, distinguished by its integration of the private key mechanism. Central to DKiS is the novel concept of decay weight, which strategically controls the transfer of information from the secret pipeline to the host pipeline. This is based on the insight that the relevance of information for the host pipeline diminishes progressively along the secret pipeline. The implementation of decay weight has notably enhanced the overall performance of the method.

A pivotal feature of DKiS is the incorporation of a private key into the steganographic process. This addition effectively addresses the longstanding challenges of balancing widespread accessibility with robust security and vulnerable to attacks in image steganography. As a result, DKiS emerges as a viable candidate for standardization in applications like image verification, where both the publicity and security of steganographic methods are paramount.

Our extensive experimental evaluations, conducted under conditions that include rounding errors, affirm the satisfactory performance of DKiS. The attack simulation further proves its security. These results demonstrate not only the method's efficacy in securely hiding images but also its potential as a versatile tool in a wide range of practical scenarios. DKiS, therefore, stands as a significant contribution to the field of image steganography, bridging the gap between public accessibility and stringent security requirements.



\begin{thebibliography}{}
	
	\bibitem[\protect\citeauthoryear{Agustsson and
		Timofte}{2017}]{agustsson2017ntire}
	Eirikur Agustsson and Radu Timofte.
	\newblock Ntire 2017 challenge on single image super-resolution: Dataset and
	study.
	\newblock In {\em Proceedings of the IEEE conference on computer vision and
		pattern recognition workshops}, pages 126--135, 2017.
	
	\bibitem[\protect\citeauthoryear{Al-Husainy and Uliyan}{2019}]{al2019secret}
	Mohammed Abbas~Fadhil Al-Husainy and Diaa~Mohammed Uliyan.
	\newblock A secret-key image steganography technique using random chain codes.
	\newblock {\em International Journal of Technology}, 10(4):731--740, 2019.
	
	\bibitem[\protect\citeauthoryear{Almazaydeh and
		Sheshadri}{2018}]{almazaydeh2018image}
	Wa'el Ibrahim~A Almazaydeh and HS~Sheshadri.
	\newblock Image steganography using a dynamic symmetric key.
	\newblock In {\em 2018 2nd International Conference on Trends in Electronics
		and Informatics (ICOEI)}, pages 1507--1513. IEEE, 2018.
	
	\bibitem[\protect\citeauthoryear{Baluja}{2017}]{Baluja2017Hiding}
	Shumeet Baluja.
	\newblock Hiding {Images} in {Plain} {Sight}: Deep {Steganography}.
	\newblock 2017.
	
	\bibitem[\protect\citeauthoryear{Baluja}{2019}]{baluja2019hiding}
	Shumeet Baluja.
	\newblock Hiding images within images.
	\newblock {\em IEEE transactions on pattern analysis and machine intelligence},
	42(7):1685--1697, 2019.
	
	\bibitem[\protect\citeauthoryear{Barni \bgroup \em et al.\egroup
	}{2001}]{barni2001improved}
	Mauro Barni, Franco Bartolini, and Alessandro Piva.
	\newblock Improved wavelet-based watermarking through pixel-wise masking.
	\newblock {\em IEEE transactions on image processing}, 10(5):783--791, 2001.
	
	\bibitem[\protect\citeauthoryear{Chan and Cheng}{2004}]{chan2004hiding}
	Chi-Kwong Chan and Lee-Ming Cheng.
	\newblock Hiding data in images by simple lsb substitution.
	\newblock {\em Pattern recognition}, 37(3):469--474, 2004.
	
	\bibitem[\protect\citeauthoryear{Duan \bgroup \em et al.\egroup
	}{2019}]{duan2019reversible}
	Xintao Duan, Kai Jia, Baoxia Li, Daidou Guo, En~Zhang, and Chuan Qin.
	\newblock Reversible image steganography scheme based on a u-net structure.
	\newblock {\em IEEE Access}, 7:9314--9323, 2019.
	
	\bibitem[\protect\citeauthoryear{Duan \bgroup \em et al.\egroup
	}{2020a}]{duan2020highx}
	Xintao Duan, Mengxiao Gou, Nao Liu, Wenxin Wang, and Chuan Qin.
	\newblock High-capacity image steganography based on improved xception.
	\newblock {\em Sensors}, 20(24):7253, 2020.
	
	\bibitem[\protect\citeauthoryear{Duan \bgroup \em et al.\egroup
	}{2020b}]{duan2020high}
	Xintao Duan, Liu Nao, Gou Mengxiao, Dongli Yue, Zimei Xie, Yuanyuan Ma, and
	Chuan Qin.
	\newblock High-capacity image steganography based on improved fc-densenet.
	\newblock {\em IEEE Access}, 8:170174--170182, 2020.
	
	\bibitem[\protect\citeauthoryear{Fridrich \bgroup \em et al.\egroup
	}{2001}]{fridrich2001detecting}
	Jessica Fridrich, Miroslav Goljan, and Rui Du.
	\newblock Detecting lsb steganography in color, and gray-scale images.
	\newblock {\em IEEE multimedia}, 8(4):22--28, 2001.
	
	\bibitem[\protect\citeauthoryear{Hawi \bgroup \em et al.\egroup
	}{2004}]{hawi2004steganalysis}
	Tariq~Al Hawi, MA~Qutayri, and Hassan Barada.
	\newblock Steganalysis attacks on stego-images using stego-signatures and
	statistical image properties.
	\newblock In {\em 2004 IEEE Region 10 Conference TENCON 2004.}, pages 104--107.
	IEEE, 2004.
	
	\bibitem[\protect\citeauthoryear{Hsu and Wu}{1999}]{hsu1999hidden}
	Chiou-Ting Hsu and Ja-Ling Wu.
	\newblock Hidden digital watermarks in images.
	\newblock {\em IEEE Transactions on image processing}, 8(1):58--68, 1999.
	
	\bibitem[\protect\citeauthoryear{Imaizumi and
		Ozawa}{2014}]{imaizumi2014multibit}
	Shoko Imaizumi and Kei Ozawa.
	\newblock Multibit embedding algorithm for steganography of palette-based
	images.
	\newblock In {\em Image and Video Technology: 6th Pacific-Rim Symposium, PSIVT
		2013, Guanajuato, Mexico, October 28-November 1, 2013. Proceedings 6}, pages
	99--110. Springer, 2014.
	
	\bibitem[\protect\citeauthoryear{Jia \bgroup \em et al.\egroup
	}{2023}]{jia2023afcihnet}
	Xingwang Jia, Huamei Xin, Lingchen Gu, Hao Wang, Jiande Sun, and Wenbo Wan.
	\newblock Afcihnet: Attention feature-constrained network for single image
	information hiding.
	\newblock {\em Engineering Applications of Artificial Intelligence},
	126:107105, 2023.
	
	\bibitem[\protect\citeauthoryear{Jing \bgroup \em et al.\egroup
	}{2021}]{jing2021hinet}
	Junpeng Jing, Xin Deng, Mai Xu, Jianyi Wang, and Zhenyu Guan.
	\newblock Hinet: Deep image hiding by invertible network.
	\newblock In {\em Proceedings of the IEEE/CVF International Conference on
		Computer Vision}, pages 4733--4742, 2021.
	
	\bibitem[\protect\citeauthoryear{Johnson and
		Jajodia}{1998}]{Johnson1998Exploring}
	Neil~F. Johnson and Sushil Jajodia.
	\newblock Exploring steganography: Seeing the unseen.
	\newblock {\em Computer}, 31(2):26--34, 2 1998.
	
	\bibitem[\protect\citeauthoryear{Kadhim \bgroup \em et al.\egroup
	}{2019}]{kadhim2019comprehensive}
	Inas~Jawad Kadhim, Prashan Premaratne, Peter~James Vial, and Brendan Halloran.
	\newblock Comprehensive survey of image steganography: Techniques, evaluations,
	and trends in future research.
	\newblock {\em Neurocomputing}, 335:299--326, 2019.
	
	\bibitem[\protect\citeauthoryear{Karim \bgroup \em et al.\egroup
	}{2011}]{karim2011new}
	SM~Masud Karim, Md~Saifur Rahman, and Md~Ismail Hossain.
	\newblock A new approach for lsb based image steganography using secret key.
	\newblock In {\em 14th international conference on computer and information
		technology (ICCIT 2011)}, pages 286--291. IEEE, 2011.
	
	\bibitem[\protect\citeauthoryear{Kessler and
		Hosmer}{2011}]{Kessler2011Overview}
	Gary~C. Kessler and Chet Hosmer.
	\newblock {\em An {Overview} of {Steganography}}, pages 51--107.
	\newblock Elsevier, 2011.
	
	\bibitem[\protect\citeauthoryear{Kweon \bgroup \em et al.\egroup
	}{2021}]{kweon2021deep}
	Hyeokjoon Kweon, Jinsun Park, Sanghyun Woo, and Donghyeon Cho.
	\newblock Deep multi-image steganography with private keys.
	\newblock {\em Electronics}, 10(16):1906, 2021.
	
	\bibitem[\protect\citeauthoryear{Lilly}{2004}]{lilly2004device}
	Glenn~M Lilly.
	\newblock Device for and method of one-way cryptographic hashing, December~7
	2004.
	\newblock US Patent 6,829,355.
	
	\bibitem[\protect\citeauthoryear{Lin \bgroup \em et al.\egroup
	}{2014}]{cocodataset}
	Tsung{-}Yi Lin, Michael Maire, Serge~J. Belongie, Lubomir~D. Bourdev, Ross~B.
	Girshick, James Hays, Pietro Perona, Deva Ramanan, Piotr Doll{'{a} }r, and
	C.~Lawrence Zitnick.
	\newblock Microsoft {COCO:} common objects in context.
	\newblock {\em CoRR}, abs/1405.0312, 2014.
	
	\bibitem[\protect\citeauthoryear{Lu \bgroup \em et al.\egroup
	}{2021}]{lu2021large}
	Shao-Ping Lu, Rong Wang, Tao Zhong, and Paul~L Rosin.
	\newblock Large-capacity image steganography based on invertible neural
	networks.
	\newblock In {\em Proceedings of the IEEE/CVF Conference on Computer Vision and
		Pattern Recognition}, pages 10816--10825, 2021.
	
	\bibitem[\protect\citeauthoryear{Luo \bgroup \em et al.\egroup
	}{2010}]{luo2010edge}
	Weiqi Luo, Fangjun Huang, and Jiwu Huang.
	\newblock Edge adaptive image steganography based on lsb matching revisited.
	\newblock {\em IEEE Transactions on information forensics and security},
	5(2):201--214, 2010.
	
	\bibitem[\protect\citeauthoryear{Nguyen \bgroup \em et al.\egroup
	}{2006}]{nguyen2006multi}
	Bui~Cong Nguyen, Sang~Moon Yoon, and Heung-Kyu Lee.
	\newblock Multi bit plane image steganography.
	\newblock In {\em Digital Watermarking: 5th International Workshop, IWDW 2006,
		Jeju Island, Korea, November 8-10, 2006. Proceedings 5}, pages 61--70.
	Springer, 2006.
	
	\bibitem[\protect\citeauthoryear{OpenAI}{2021}]{chatgpt}
	OpenAI.
	\newblock Chatgpt: A language model for conversational agents.
	\newblock {\em OpenAI Blog}, 2021.
	
	\bibitem[\protect\citeauthoryear{Pan \bgroup \em et al.\egroup
	}{2011}]{pan2011image}
	Feng Pan, Jun Li, and Xiaoyuan Yang.
	\newblock Image steganography method based on pvd and modulus function.
	\newblock In {\em 2011 International Conference on Electronics, Communications
		and Control (ICECC)}, pages 282--284. IEEE, 2011.
	
	\bibitem[\protect\citeauthoryear{Ruanaidh \bgroup \em et al.\egroup
	}{1996}]{ruanaidh1996phase}
	JJKO Ruanaidh, William~J Dowling, and Francis~M Boland.
	\newblock Phase watermarking of digital images.
	\newblock In {\em Proceedings of 3rd IEEE International Conference on Image
		Processing}, volume~3, pages 239--242. IEEE, 1996.
	
	\bibitem[\protect\citeauthoryear{Russakovsky \bgroup \em et al.\egroup
	}{2015}]{ILSVRC15}
	Olga Russakovsky, Jia Deng, Hao Su, Jonathan Krause, Sanjeev Satheesh, Sean Ma,
	Zhiheng Huang, Andrej Karpathy, Aditya Khosla, Michael Bernstein,
	Alexander~C. Berg, and Li~Fei-Fei.
	\newblock {ImageNet Large Scale Visual Recognition Challenge}.
	\newblock {\em International Journal of Computer Vision (IJCV)},
	115(3):211--252, 2015.
	
	\bibitem[\protect\citeauthoryear{Setiadi}{2021}]{setiadi2021psnr}
	De~Rosal Igantius~Moses Setiadi.
	\newblock Psnr vs ssim: imperceptibility quality assessment for image
	steganography.
	\newblock {\em Multimedia Tools and Applications}, 80(6):8423--8444, 2021.
	
	\bibitem[\protect\citeauthoryear{Shi \bgroup \em et al.\egroup
	}{2018}]{shi2018ssgan}
	Haichao Shi, Jing Dong, Wei Wang, Yinlong Qian, and Xiaoyu Zhang.
	\newblock Ssgan: Secure steganography based on generative adversarial networks.
	\newblock In {\em Advances in Multimedia Information Processing--PCM 2017: 18th
		Pacific-Rim Conference on Multimedia, Harbin, China, September 28-29, 2017,
		Revised Selected Papers, Part I 18}, pages 534--544. Springer, 2018.
	
	\bibitem[\protect\citeauthoryear{Tamimi \bgroup \em et al.\egroup
	}{2013}]{tamimi2013hiding}
	Abdelfatah~A Tamimi, Ayman~M Abdalla, and Omaima Al-Allaf.
	\newblock Hiding an image inside another image using variable-rate
	steganography.
	\newblock {\em International Journal of Advanced Computer Science and
		Applications (IJACSA)}, 4(10), 2013.
	
	\bibitem[\protect\citeauthoryear{Tsai \bgroup \em et al.\egroup
	}{2009}]{tsai2009reversible}
	Piyu Tsai, Yu-Chen Hu, and Hsiu-Lien Yeh.
	\newblock Reversible image hiding scheme using predictive coding and histogram
	shifting.
	\newblock {\em Signal processing}, 89(6):1129--1143, 2009.
	
	\bibitem[\protect\citeauthoryear{Xiao \bgroup \em et al.\egroup
	}{2020}]{xiao2020invertible}
	Mingqing Xiao, Shuxin Zheng, Chang Liu, Yaolong Wang, Di~He, Guolin Ke, Jiang
	Bian, Zhouchen Lin, and Tie-Yan Liu.
	\newblock Invertible image rescaling.
	\newblock In {\em Computer Vision--ECCV 2020: 16th European Conference,
		Glasgow, UK, August 23--28, 2020, Proceedings, Part I 16}, pages 126--144.
	Springer, 2020.
	
	\bibitem[\protect\citeauthoryear{Xu \bgroup \em et al.\egroup }{2022}]{RIIS}
	Youmin Xu, Chong Mou, Yujie Hu, Jingfen Xie, and Jian Zhang.
	\newblock Robust invertible image steganography.
	\newblock In {\em Proceedings of the IEEE/CVF Conference on Computer Vision and
		Pattern Recognition}, pages 7875--7884, 2022.
	
	\bibitem[\protect\citeauthoryear{Yang \bgroup \em et al.\egroup
	}{2023}]{yang2023pris}
	Hang Yang, Yitian Xu, Xuhua Liu, and Xiaodong Ma.
	\newblock Pris: Practical robust invertible network for image steganography.
	\newblock {\em arXiv preprint arXiv:2309.13620}, 2023.
	
	\bibitem[\protect\citeauthoryear{Zhang \bgroup \em et al.\egroup
	}{2019}]{zhang2019steganogan}
	Kevin~Alex Zhang, Alfredo Cuesta-Infante, Lei Xu, and Kalyan Veeramachaneni.
	\newblock Steganogan: High capacity image steganography with gans.
	\newblock {\em arXiv preprint arXiv:1901.03892}, 2019.
	
	\bibitem[\protect\citeauthoryear{Zhang \bgroup \em et al.\egroup
	}{2020}]{zhang2020udh}
	Chaoning Zhang, Philipp Benz, Adil Karjauv, Geng Sun, and In~So Kweon.
	\newblock Udh: Universal deep hiding for steganography, watermarking, and light
	field messaging.
	\newblock {\em Advances in Neural Information Processing Systems},
	33:10223--10234, 2020.
	
	\bibitem[\protect\citeauthoryear{Zhi \bgroup \em et al.\egroup
	}{2003}]{zhi2003lsb}
	Li~Zhi, Sui~Ai Fen, and Yang~Yi Xian.
	\newblock A lsb steganography detection algorithm.
	\newblock In {\em 14th IEEE Proceedings on Personal, Indoor and Mobile Radio
		Communications, 2003. PIMRC 2003.}, volume~3, pages 2780--2783. IEEE, 2003.
	
	\bibitem[\protect\citeauthoryear{Zhong \bgroup \em et al.\egroup
	}{2019}]{zhong2019publaynet}
	Xu~Zhong, Jianbin Tang, and Antonio~Jimeno Yepes.
	\newblock Publaynet: largest dataset ever for document layout analysis.
	\newblock In {\em 2019 International Conference on Document Analysis and
		Recognition (ICDAR)}, pages 1015--1022. IEEE, Sep. 2019.
	
	\bibitem[\protect\citeauthoryear{Zhu \bgroup \em et al.\egroup
	}{2018}]{zhu2018hidden}
	Jiren Zhu, Russell Kaplan, Justin Johnson, and Li~Fei-Fei.
	\newblock Hidden: Hiding data with deep networks.
	\newblock In {\em Proceedings of the European conference on computer vision
		(ECCV)}, pages 657--672, 2018.
	
\end{thebibliography}

\end{document}